\documentclass[12pt,epsf]{article}
\usepackage{graphicx, amsmath}

\begin{document}
\sloppy
\begin{flushright}{SIT-HEP/TM-60}
\end{flushright}
\vskip 1.5 truecm
\centerline{\large{\bf Casimir force for cosmological domain walls}} 
\vskip .75 truecm
\centerline{\bf Tomohiro Matsuda\footnote{matsuda@sit.ac.jp}
\footnote{On sabbatical leave of absence from Saitama institute of
technology for the academic year 2011-2012. 
This paper has been submitted from Nagoya university and revised at
Lancaster university.}}
\vskip .4 truecm
\centerline {\it Laboratory of Physics, Saitama Institute of Technology,
Fukaya, Saitama 369-0293, Japan}
\centerline{\it Department of Physics, Nagoya University, Nagoya
464-8602, Japan}  
\centerline {\it Physics Department, Lancaster University,
 Lancaster LA1 4YB, U.K.}
\vskip 1. truecm

\makeatletter
\@addtoreset{equation}{section}
\def\theequation{\thesection.\arabic{equation}}
\makeatother
\vskip 1. truecm

\begin{abstract}
\hspace*{\parindent}
We calculate the vacuum fluctuations that may affect the evolution of
cosmological domain walls.
Considering domain walls, which are classically stable and have
 interaction with a scalar field, we  show that explicit symmetry
 violation in the interaction may cause quantum bias that can solve the
 cosmological domain wall problem.
\end{abstract}

\newpage
\section{Introduction}

The Casimir effect suggested originally in 1948 has been used to
understand the contribution from the vacuum fluctuations of quantum
fields\cite{Casimir:1948dh}.
The variation in the vacuum fluctuations, which appears in the
excitations as the consequence of the non-trivial boundary conditions or the
topology of the space, causes a shift of the vacuum energy.
The original model, which appears as the two conducting parallel plates
in the free $R^3$ space, the attractive force is confirmed
experimentally by Sparnaay\cite{Sparnaay:1958wg} and a more precise result
have been given more recently in Ref.\cite{Bressi:2002fr}.

For the simplest one-dimensional model, the sum of excitations in the 
 (1+1)-dimensional spacetime, where the boundaries are separated by the
 distance $L$, is  given by 
\begin{equation}
E_L\equiv \sum^{\infty}_{n=1} \frac{1}{2}\hbar  
\omega_n\equiv \frac{\pi\hbar}{2L}\sum^{\infty}_{n=1}n,
\end{equation}
where $\hbar$ is the reduced Planck constant.
We define the averaged energy density by
\begin{equation}
\rho_L\equiv \frac{E_L}{L} =
 \frac{\pi\hbar}{2L^2}\sum^{\infty}_{n=1}n.
\end{equation}
Hereafter we set $\hbar=c=1$. 
In the limit of $L\rightarrow \infty$, this gives for the massless field
\begin{equation}
\rho_{\infty}\equiv \frac{E_\infty}{L}= \int_0^{\infty} \frac{dk}{2\pi} k.
\end{equation}
where $k$ denotes the continuous ($\L\rightarrow \infty$) limit of
$k_n \equiv \frac{\pi}{L}n$.
The Casimir energy is defined (regularized) by $\Delta E\equiv E_L-E_{\infty}$.
To obtain a finite result, consider the regularization\cite{curved-book}
\begin{equation}
\hat{\rho}_L\equiv \frac{1}{2L}\sum^{\infty}_{n=1} \omega_n
 e^{-\omega_n/\Lambda}
=\frac{\Lambda^2}{2\pi}-\frac{\pi}{24L^2}+{\cal O}(\frac{1}{\Lambda})
\end{equation}
and 
\begin{equation}
\hat{\rho}_\infty\equiv 
\frac{1}{2\pi}\int_0^\infty d\omega w e^{-\omega/\Lambda}
=\frac{\Lambda^2}{2\pi},
\end{equation}
where $\Lambda$ is introduced as the manifestation of the cut-off scale. 
Here $e^{-\omega_n/\Lambda}\rightarrow 0$ is assumed for $n\rightarrow \infty$.
Regularization using $\zeta$-function is also possible.
Considering the regularization, the energy shift caused by the
boundary is estimated as
\begin{equation}
\rho^R \equiv \rho_L-\rho_\infty \simeq -\frac{\pi}{24L^2}.
\end{equation}

\underline{Domain Walls in cosmology}

In the context of the hot big bang theory, the fundamental theory of
unification predicts a sequence of phase transitions during the
cosmological evolution of the Universe.
These phase transitions can be accompanied by the formation of domain
structures that is determined by the symmetry breaking at the phase
transition.
Wall domination, which always leads to a serious problem if the energy
scale of the domain wall exceed 1MeV, can be avoided if a small
bias $\delta \rho \equiv \epsilon\ne 0$ appears.
The bias (here we use this word specifically for the energy
difference between the false and the true vacua) becomes important when
the force per unit area on the walls becomes comparable to the tension
of the wall. 
Then, the condition for the successful decay of the cosmological domain
walls is satisfied when\cite{Vilenkin-book}
\begin{equation}
\epsilon > G \sigma^2,
\end{equation}
where $\sigma$ denotes the tension of the wall.

The condition shows that the global discrete symmetry, which leads to the
domain wall formation, must be broken explicitly. 
In that case, the magnitude of the (explicit) breaking parameter must
explain the bias $\epsilon  > G \sigma^2$.
This idea is useful in supersymmetric theory, in which the
supergravity potential breaks discrete symmetry with the required
magnitude\cite{matsuda-wall}.
Usually, the origin of the bias is considered for the explicit symmetry
breaking in the potential.
Walls that are formed after brane inflation is realized by the
deformation of the brane configuration in the compactified
space\cite{brane-wall}.  
Wall-like structure observed on a cosmic string may appear as a monopoles
connected by the strings, which turns out to be a 
so-called cosmic necklace\cite{Vilenkin-book, matsuda-necklaces}

In this paper, we consider a simple model in which the mass of an
additional field $\chi$ is induced by the interaction
\begin{equation}
{\cal L}_{int}=\frac{1}{2}g^2 \phi^2 \chi^2,
\end{equation}
where $\phi$ is the field that forms the wall configuration.
It would be easy to find that an explicit symmetry breaking in the
interaction causes a small mass difference in $m_\chi$.
The small mass difference between the adjacent vacua, which may be very small
compared with the energy scale of the domain wall, may cause a 
bias when the vacuum fluctuations are considered.
In addition to the conventional vacuum fluctuations, which may be the
dominant contribution, the mass difference causes
a boundary for the excitations of the $\chi$-field, which leads to
another source of the bias.
Calculation of the vacuum fluctuations (i.e., the quantum bias caused by
the Casimir effect) shows that the Casimir force may have a significant
impact on the evolution of the cosmological domain walls.

Although the Casimir force for the massless field seems to be consistent
with the experiments, it may have serious drawbacks\footnote{The
regularization of the Casimir energy is under control. On the
other hand, the root of the cosmological constant problem is not quite
obvious. At this moment, it is not obvious whether an improvement is
required in the regularization in solving the cosmological constant
problem. The word ``drawbacks'' has been used to mention the specific
situation in relation to the cosmological constant.}  
 when it is applied to the issue of the cosmological constant.
Since we are considering quantum bias between adjacent vacua, we cannot
be free from the peculiar assumptions that are needed to understand or
explain the (almost) vanishing cosmological constant. 
In this paper we are making best effort to find a sensible result for
the quantum bias, but it should be noted that our results are based
on these assumptions, which will be further explained in Sec.\ref{sec3}.

\section{Casimir energy for the domain walls with a small mass-gap (1+1
 dimensional toy model)}

\begin{figure}[h]
 \begin{center}
\begin{picture}(200,240)(100,0)
\resizebox{12cm}{!}{\includegraphics{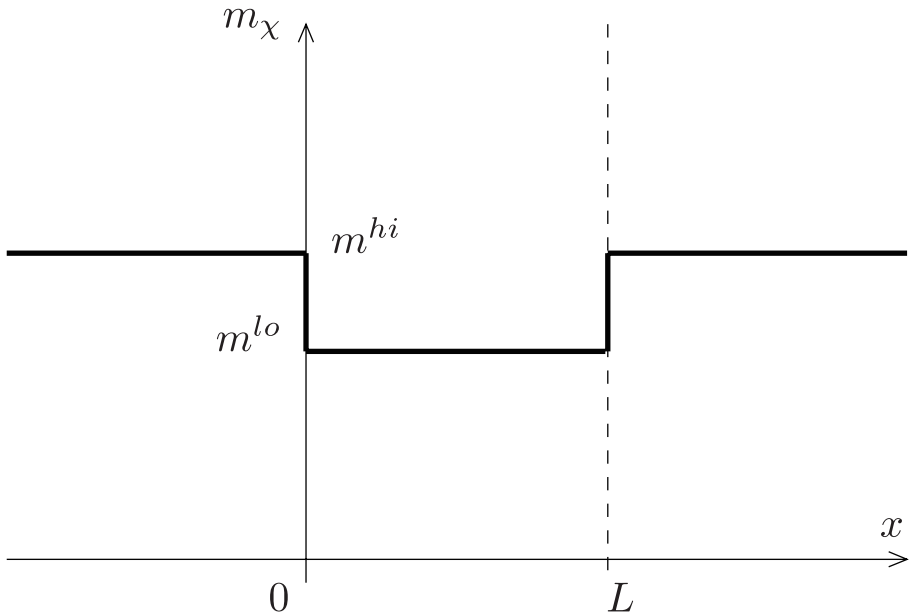}} 
\end{picture}
 \caption{The field $\chi$ feels a mass-gap at the domain walls, which
  are placed at $x=0$ and $x=L$. The excitations that are trapped inside
  $0<x<L$ are discrete.}     
\label{fig:gap}
 \end{center}
\end{figure}

In the four dimensional model, it is possible to consider a double-well
potential 
$V(\phi)=\frac{1}{4}\lambda \left(\phi^2-v^2\right)^2$, which has a
$Z_2$ symmetry ($\phi=+v \leftrightarrow \phi=-v$).
This symmetry is broken spontaneously in either vacuum.
Interaction with a real scalar field $\chi$ can be given by ${\cal
L}_{int}=\frac{1}{2} g \phi^2 \chi^2$, which gives a mass to the field $\chi$. 
The domain walls are kinks of a real scalar field $\phi$, which mediate
between the two vacua $\phi=v$ and $\phi=-v$.
As we mentioned, we do not consider explicit breaking of the discrete
symmetry in the potential $V(\phi)$.
Instead, we add a small but explicit symmetry breaking to the
interaction, so that it leads to a small mass-gap for the $\chi$-field.
Obviously, there is no bias in the classical vacua.
The source of the quantum bias can be explained by a small breaking of
the $Z_2$ symmetry in the interaction, which can be expressed as
\begin{equation} 
{\cal L}_{int}=\frac{1}{2} g^2 (\phi-\epsilon_z)^2 \chi^2,
\end{equation}
where $\epsilon_z$ measures the explicit symmetry breaking in the
interaction.\footnote{We pointed out that the bias
introduced by the Casimir effect may be important for the evolution of
the cosmological domain walls. On the other hand, if one diagonalizes the whole
Lagrangian, although the calculation is highly model-dependent and is
not suitable for our argument, these domain walls may be unstable
classically. In that case 
the bias introduced by the Casimir effect may be smaller than the
classical bias. Note that we are not arguing that the classical
bias is always smaller than the Casimir effect.}
Denoting the mass of $\chi$ in each domain by
$m_{\chi}^{lo}\equiv g(v-\epsilon_z)$ and $m_{\chi}^{hi}\equiv g(
v+\epsilon_z)$, and placing the lower-mass domain in the area sandwiched
by two domain walls, the $\chi$-field excitations
that are discretized (trapped) by the boundary have 
 $\omega_n ^2 = k_n^2+(m^{lo})^2 < (m^{hi})^2$.
Obviously, the excitations with $\omega_n< m_\chi^{hi}$ can exist
only in the domain sandwiched by the walls, while other (higher)
excitations are continuous in both domains.\footnote{See fig.\ref{fig:gap}.} 

Let us first consider a simple model in one dimensional space ($x$), and
place domain walls at $x=0$ and $x=L$. 
The sum of the excitations discretized by the domain walls leads to the
energy density 
\begin{equation}
\rho_{trap}\equiv \frac{1}{2L}\sum_{n=1}^{n_{Max}} \omega_n,
\end{equation}
where the integer $n_{Max}$ is approximately given by $n_{Max}\sim
\frac{L}{\pi}\sqrt{(m_\chi^{hi})^2 -(m_\chi^{lo})^2}=\frac{2Lg}{\pi}
\sqrt{v\epsilon_z}$.
For the estimation of the Casimir effect,
we set $n_{Max}=\frac{2Lg}{\pi} \sqrt{v\epsilon_z}$ hereafter.
With regard to the wavelength of these excitations, the effective length
$L^{eff}$ may depend on $\omega$.
The difference $L^{eff}(\omega)-L\ne 0$ may be significant near $n_{Max}$,
where the excitations penetrate into the higher-mass domains.
However, for the simple estimation of the Casimir effect, we choose the
approximation $L^{eff}(\omega)=L$ in this paper.

Despite the simplicity of the scenario, regularization of the vacuum
fluctuations requires non-trivial assumptions that are far from obvious.
In order to compare our result with the usual Casimir effect, it would be
useful to start with a massless field.
Therefore, we first consider a
model with $m^{lo}=0$ and $m^{hi}\ne 0$, so that we can calculate the 
Casimir energy using the conventional assumptions.
The Casimir energy density in the constrained massless domain,
which is sandwiched by the walls at $x=0$ and $x=L$, is given by 
\begin{eqnarray}
\rho^R&\equiv& \rho_L-\rho_\infty=\frac{1}{2L}\sum_{n=1}^{n_{Max}}
 \frac{\pi}{L}n 
-\int^{k_{Max}}_0 \frac{dk}{2\pi} k,
\end{eqnarray}
where the cancellation occurs in the continuous part $k>k_{Max}$.
Here we set $n_{Max}=\frac{Lm^{hi}}{\pi}$ and $k_{Max}=m^{hi}$.
Executing the finite sum, it leads to 
\begin{equation}
\rho^R=
\frac{\pi}{2L^2}\frac{n_{Max}(n_{Max}-1)}{2} - \frac{1}{4\pi}(m^{hi})^2
= -\frac{1}{4L}m^{hi},
\end{equation}
where $n_{Max}=\frac{Lm^{hi}}{\pi}> 1$ is required to obtain a non-trivial result.

In the above calculation, the origin of the Casimir energy is 
the discretization of the excitations, which is obviously finite.
We subtracted the ``common'' part, which is continuous and divergent.  
In doing this, we assumed that the discretization of the excitations
does not affect the regularization of the continuous part above
$k>k_{Max}$.  

In the above calculation, we defined the Casimir energy in the massless
domain by $\rho_L-\rho_\infty$\cite{curved-book}. 
However, the bias between the adjacent vacua may ``not'' be measured by the
Casimir effect in the massless domain.
Namely, there is the possibility that $\rho^{lo}_\infty-\rho^{hi}_\infty$ may
become the dominant part of the bias.  
In such calculation we have to reconsider regularization in the massive domain.

To understand the problem, consider 
two domains denoted by ``A'' and ``B'', which are separated by
a wall.
Then the vacuum fluctuations are (naively) given by
$\rho^{A}_\infty \equiv \int_0^\infty \frac{dk}{2\pi} \sqrt{k^2+m_A^2}$
and $\rho^{B}_\infty \equiv \int_0^\infty \frac{dk}{2\pi}
\sqrt{k^2+m_B^2}$, respectively.
Without additional principle for the regularization, $m_A\ne m_B$
leads to $|\rho^A_\infty-\rho^B_\infty|\sim m_A^2-m_B^2$.
If the above calculation is true, the total vacuum fluctuations 
may depend explicitly on the particle content of the vacuum.
The simplest way to avoid this problem is to assume some (unknown)
regularization scheme that explains vanishing vacuum fluctuations in the
$L\rightarrow\infty$ limit, which works for each field.
If this assumption is true, one always find $\rho^i_\infty=0$ for the
entire field
labeled by $i$. 
Instead, one may consider a delicate cancellation between
fields with different masses and spins, which eventually leads to
the effective cosmological constant $\Lambda_c^{eff}\equiv \Lambda_0 +
\sum \rho^i_\infty\simeq0$, where $\Lambda_0$ denotes contributions from other
effects.   
Obviously, in the latter case the delicate cancellation is crucial for the bias
calculation. 
Namely, if the delicate cancellation is violated in a false-vacuum domain, where
the mass distribution is different from the true vacuum, 
the quantum bias is more significant compared with the Casimir energy
calculated above.

In the next section, we consider a realistic four-dimensional model of
cosmological domain walls and examine the cosmological domain wall
problem with the use of the quantum bias, which is caused by the symmetry
breaking in the interaction.

\section{Casimir energy for the domain walls with a small mass-gap}
\label{sec3}
We first consider a massless domain sandwiched by massive domains.
For this thought experiment, specific form of the potential can be
expressed by $V(\phi)\sim \phi^2(\phi-v)^2$ (in this case there is no
$Z_2$ symmetry but the degeneracy of the classical vacua still remains), or
alternatively a fine-tuning can be considered
in the interaction term (i.e., $v=\epsilon_z$ makes $m^{lo}=0$).
In order to realize a small mass-gap compared with the domain-wall
tension ($m^{hi}\ll v$),  we consider $g\ll 1$.
Consider the domain sandwiched by two infinite and flat walls in the
free $R^3$ space. 
The walls are placed at distance $L$ apart and lie in the xy-plane.
The standing waves are 
\begin{equation}
\chi_n (x,y,x,t)=e^{-i \omega_n t} \left[
e^{ik_x x}e^{ik_y y}
\right] \sin k_n z,
\end{equation}
where $k_x$ and $k_y$ are the wave vectors in the direction parallel to
the walls, which are continuous, while the discretized wave vector
\begin{equation}
k_n = \frac{\pi}{L}n
\end{equation}
is  perpendicular to the walls.
Due to the ``shallow'' potential, discretization occurs for the
low energy excitations $k\le k_{Max}$.
The vacuum fluctuations in the massless domain is given by
\begin{equation}
\rho_L =\int \frac{dk_x dk_y}{(2\pi)^2}
\left[\frac{1}{2L}\sum^{n_{max}}_{n=1} \omega_n +
 \int_{k_{max}}^{\infty} \frac{dk_z}{(2\pi)}\omega 
\right],
\end{equation}
where the discretized wave ($\omega_n$) is given by
\begin{equation}
\omega_n \equiv \left[k_x^2+k_y^2 + \left(\frac{n \pi}{L}\right)^2
\right]^{1/2},
\end{equation}
while for the continuous wave, it is given by
\begin{equation}
\omega \equiv \left[k_x^2+k_y^2 + k_z^2 \right]^{1/2}.
\end{equation}
After integration and subtraction of $\rho_\infty$, it leads to the
regularized vacuum energy\cite{curved-book}
\begin{eqnarray}
\rho^R &\equiv& \rho_L-\rho_\infty\nonumber\\
&=& \frac{1}{12\pi}\left[
- \frac{\pi^3}{2L^4}\sum_{n=1}^{n_{max}}n^3+
\int^{k_{max}}_0\frac{dk_z}{2\pi}k_z^3,
\right]\nonumber\\
&=&  \frac{1}{12\pi}\left[
-\frac{\pi^3}{2L^4}\frac{n^2_{max}(n_{max}+1)^2}{4}+
\frac{k_{max}^4}{2\pi},
\right]\nonumber\\
&=&  -\frac{1}{96}\left[\frac{2m^3}{L}+\frac{\pi m^2}{L^2}
\right],
\end{eqnarray}
where the last equation is derived for
$k_{max}=\frac{\pi}{L}n_{max}\equiv m$. Here $m$ denotes the mass of the
field $\chi$ in the massive domain.

In the above calculation, we considered a massless domain sandwiched by
massive domains with a mass-gap $\delta m_\chi=m$ at the boundary
(domain wall).  
Unlike the conventional Casimir effect, however, the discretization
occurs only for a finite number of the excitations.

The above calculation for the ``massless'' domain is straight and
very instructive, but here we must remember that the most important 
situation in our model is that the ``massive'' field feels a small
mass-gap at a domain wall.  
In that case, the Casimir energy may cause the energy difference 
that depends on the mass distribution, as we already mentioned in the 
previous section for the domain walls.
The situation seems unfavorable for the cosmological constant
problem.
Since the root of the cosmological constant problem is not
obvious yet, and this is not the topic I am discussing in this paper, 
we follow the conventional calculation and try to find sensible
consequences that may have phenomenological interest. 

For instance, let us consider a bias $\epsilon$ for the cosmological
domain walls with the tension $\sigma \sim v^3$.
If the (almost) vanishing cosmological constant is explained by the
delicate cancellation ($\Lambda_c^{eff}\equiv \Lambda_0 +
\sum \rho^i_\infty\simeq0$), the bias between the domains with
$m_A=m+\delta m$ and $m_B=m$ is calculated as 
\begin{equation}
\epsilon \sim (m+\delta m)^4 -m^4\sim 4m^3 \delta m.
\end{equation}
For the interaction that leads to $m = gv$, the required condition for
the safe decay is 
\begin{equation}
\delta m > \delta_c \equiv\frac{v^3}{4M_p^2 g^3} =
\frac{1}{4g^3} \left(\frac{v}{M_p}\right)^2 v,
\end{equation}
which is about $g^{-3}(v/M_p)^2 \ll 1$ times smaller than
$v$.\footnote{Even if $g$ is very small, the condition is conceivable for
$v\sim$TeV $\ll M_p$ domain walls. } 
In this case, we may conclude that a small mass-gap that is
caused by the symmetry breaking in the interaction term
can be used to solve the cosmological domain walls problem.
We find, therefore, a very useful solution to the
cosmological domain-wall problem when the cosmological constant is tuned
by the delicate cancellation.

If one cannot agree with such cancellation, an alternative
assumption would be that the Casimir energy vanishes for each massive
field in the infinite-volume limit (i.e., $\rho^i_\infty=0$ for any $i$).
Our purpose in this paper is not to argue which assumption is
plausible, but to address the consequences that result from these 
assumptions.
If the result obtained from our calculation turns out to be false in some
experiment, one needs to introduce an additional principle for the
regularization, so that one can calculate correctly the Casimir effect
caused by the mass-gap.
In any case, we believe that studying cosmological domain walls
in terms of the Casimir effect can make a difference to
the usual approaches to these problems.

Note however that, as far as the ``Casimir energy'' is defined using
$\rho_L-\rho_\infty$,  it is always possible to calculate the 
Casimir energy in an automatic manner. 
Therefore, in this paragraph, we are not going to argue the authenticity of
the Casimir effect for a massive field, which may need improvement if it
is responsible for the cosmological constant, but to calculate the Casimir
effect in the automatic way. 
Consider two domains in which the masses of the field $\chi$ are given by
$(m^{hi})^2=m^2+ \delta m^2$  in the high-mass domain and $m^{lo}=m$
in the low-mass domain.\footnote{In this notation $\delta m^2$ is not
identical to $(\delta m)^2$.}
The vacuum fluctuations in the low-mass domain, which is sandwiched by the two
walls and the high-mass domains outside, is given by
\begin{equation}
\rho_L =\frac{1}{L}\int \frac{dk_x dk_y}{(2\pi)^2}
\left[\frac{1}{2}\sum^{n_{max}}_{n=1} \omega_n + \int_{k_{max}}^{\infty} \frac{dk_z}{(2\pi)}\omega
\right].
\end{equation}
Here, for the discretized wave, $\omega_n$  is given by
\begin{equation}
\omega_n \equiv \left[m^2+k_x^2+k_y^2 + \left(\frac{n \pi}{L}\right)^2
\right]^{1/2},
\end{equation}
while for the continuous wave, $\omega$ is given by
\begin{equation}
\omega \equiv \left[m^2+k_x^2+k_y^2 + k_z^2 \right]^{1/2}.
\end{equation}
After integration and subtraction of $\rho_\infty$, it leads to the
regularized vacuum energy\footnote{Here the word ``regularization''
means specifically the subtraction of $\rho_\infty$. See also
ref.\cite{curved-book} in which ``normal ordering'' has been discussed
 in relation to the regularization.}
\begin{eqnarray}
\rho^R &\equiv& \rho_L-\rho_\infty\nonumber\\
&=& \frac{1}{12\pi}\left[
- \frac{1}{2L}\sum_{n=1}^{n_{max}}\left[m^2+\left(\frac{n\pi}{L}\right)^2
\right]^{3/2}+
\int^{k_{max}}_0\frac{dk_z}{2\pi}\left(m^2+k_z^2\right)^{3/2}
\right],
\end{eqnarray}
where we set $k_{max}\simeq \frac{\pi}{L}n_{max}\equiv \sqrt{\delta m^2}$.
In the limit of  $\delta m^2/m^2 \ll 1$ we consider the approximation
\begin{equation}
\left(m^2+k_z^2\right)^{3/2}\simeq m^3 + \frac{3}{2}m k_z^2,
\end{equation}
which leads to 
\begin{eqnarray}
\rho^R &\simeq& 
- \frac{1}{24\pi L}\left[n_{Max}m^3 + \frac{\pi^2 m}{4L^2}
n_{Max}(n_{Max}+1)(2n_{Max}+1)\right]
+\frac{1}{24\pi^2}m^3 k_{Max}+\frac{1}{48\pi^2}m k_{Max}^3\nonumber\\
&=& -\frac{1}{96}\left[\frac{3m \delta m^2}{\pi
		  L}+\frac{m}{L^2}\sqrt{\delta m^2} 
\right].
\end{eqnarray}

\begin{figure}[h]
 \begin{center}
\begin{picture}(200,80)(100,0)
\resizebox{15cm}{!}{\includegraphics{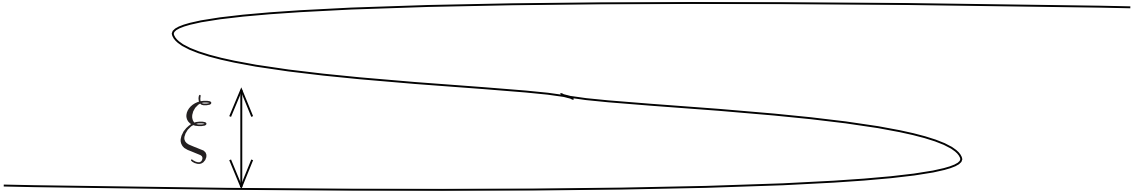}} 
\end{picture}
 \caption{Wiggles or foldings may appear in the small-scale structure of
  the cosmological domain wall.}    
\label{fig:folded-wall}
 \end{center}
\end{figure}

Going back to cosmological domain walls, we find immediately  that
the walls may appear equally in the $x$ and $y$ directions.
Also, the shape of the domain may affect the calculation.
However, a simple estimation of the vacuum fluctuation is not difficult,
which leads to
\begin{equation}
\rho^R \sim c\frac{m \delta m^2}{\xi(t)},
\end{equation}
where $c\sim 10^{-2}$ is a numerical constant and $\xi(t)$ denotes the
distance between walls. 
Since the Hubble parameter at the beginning of the wall domination is
$H_d\simeq \sigma/M_p^2$\cite{Vilenkin-book},
the Casimir energy at the wall domination is expressed as
\begin{equation}
\rho^R \sim c\frac{\sigma}{M_p^2}m\delta m^2,
\end{equation}
where the typical scale of the wall structure is assumed to be
$\xi_d \sim H^{-1}_d$. 
Considering $\delta m^2 \equiv (m+\delta m)^2-m^2 = 2m \delta
m+(\delta m)^2$ in our notation, the Casimir force may satisfy the bias
condition $\epsilon_z > \sigma^2/M_p^2$ when 
\begin{equation}
\delta m> \frac{v}{cg^2}.
\end{equation}
Therefore, the ``Casimir force'' calculated above seems to be
unimportant for the domain wall problem.
On the other hand, structures like wiggles or foldings may
typically have much smaller scale compared with $H^{-1}$.
The evolution of these small-scale structures of the cosmological domain
walls may be
affected by the Casimir force.\footnote{See fig.\ref{fig:folded-wall}}.

\underline{Can the multiplicity enhances the Casimir effect?}

If the field $\chi$ has the origin in some higher dimensional theory,
one cannot neglect the multiplicity of the field.
Namely, if the explicit symmetry breaking in the interaction term
causes a mass-gap $\sim \delta m$ to the lowest state,
it may also cause the same mass-gap to the Kaluza-Klein
states. 
Then, the Casimir force calculated above can be enhanced by the number
of the Kaluza-Klein states.
For the specific scenario, we consider the mass for the k-th Kaluza-Klein
state;
\begin{equation}
m_{k}^2 = m^2 + \frac{k^2}{R_5^2},
\end{equation}
where $k$ is an integer.
Then the Casimir energy is calculated as 
\begin{eqnarray}
\rho^R &\equiv& \rho_L-\rho_\infty\nonumber\\
&=& \frac{1}{12\pi}\sum_k\left[
- \frac{1}{2L}\sum_{n=1}^{n_{max}}\left[m_k^2+\left(\frac{n\pi}{L}\right)^2
\right]^{3/2}+
\int^{k_{max}}_0\frac{dk_z}{2\pi}\left(m_k^2+k_z^2\right)^{3/2}
\right]\nonumber\\
&\simeq& -\frac{1}{96}
\left[\frac{3(\delta m^2)}{\pi L}+\frac{\delta m}{L^2}
\right]\sum_k\sqrt{m^2 + \frac{k^2}{R_5^2}},
\end{eqnarray}
where the result can be expressed formally in terms of Epstein
$\zeta$-function.\footnote{The Casimir
energy in relation to extra dimensions has been studied by many
authors\cite{Casimir-KK, Casimir-KK-uni}. 
With regard to brane models, the finite temperature Casimir force due to
a massless scalar in the bulk of a brane model has been calculated in
ref.\cite{Casimir-KK-T}.   
The Casimir energy of massless and massive bulk fields can generate
 a potential that stabilizes the radius of the compact direction while
 it may driving the accelerated expansion in the non-compact
directions\cite{Casimir-KK-mixed-stable}.
The Casimir force acting on two parallel planes lying within the single
brane of a Randall-Sundrum scenario has been discussed in
Ref.\cite{Casimir-KK-warped}. }

\section{Conclusions}
In this paper we considered two types of vacuum fluctuations for the
evolution of the cosmological domain walls.
We considered a potential which does ``not'' break explicitly the $Z_2$
symmetry.
Instead, we added the interaction that breaks explicitly the
symmetry.
This term does not cause any bias in the classical vacua, but may source
the quantum bias.
Then the two vacua at $\phi=\pm v$, which are classically degenerate,
can be split by the Casimir effect.
In the first example, in which we compared the vacuum fluctuations in
different domains assuming that the vanishing cosmological
constant is explained by the delicate cancellation,
we find that domain walls can decay safe due to the quantum bias.
In the second example, in which we considered the effect of the
boundaries that are formed by the domain walls, the mass gap leads to a
discretization of the excitations.  
The latter effect may be important for the small-scale structures of the domain
walls, while it may  be unimportant for the safe decay.

\section{Acknowledgment}
We wish to thank K.Shima for encouragement, and our colleagues at
Nagoya university and Lancaster university for their kind hospitality.


\begin{thebibliography}{1}
\bibitem{Casimir:1948dh}
  H.~B.~G.~Casimir,
  ``On the Attraction Between Two Perfectly Conducting Plates,''
  Indag.\ Math.\  {\bf 10}, 261 (1948)
  [Kon.\ Ned.\ Akad.\ Wetensch.\ Proc.\  {\bf 51}, 793 (1948\
	FRPHA,65,342-344.1987\ KNAWA,100N3-4,61-63.1997)]. 
\bibitem{Sparnaay:1958wg}
  M.~J.~Sparnaay,
  ``Measurements of attractive forces between flat plates,''
  Physica {\bf 24}, 751 (1958).
\bibitem{Bressi:2002fr}
  G.~Bressi, G.~Carugno, R.~Onofrio and G.~Ruoso,
  ``Measurement of the Casimir force between parallel metallic surfaces,''
  Phys.\ Rev.\ Lett.\  {\bf 88}, 041804 (2002)
  [arXiv:quant-ph/0203002].
\bibitem{curved-book}
  N.~D.~Birrell, P.~C.~W.~Davies,
  ``Quantum Fields In Curved Space,''
  Cambridge, Uk: Univ. Pr. ( 1982) 340p.  
\bibitem{Vilenkin-book}
A. ~Vilenkin and E.~P.~S.~Shallard, 
``COSMIC STRINGS AND OTHER TOPOLOGICAL DEFECTS'', Cambridge Univ. Press.
\bibitem{matsuda-wall}
  T.~Matsuda,
  ``Weak scale inflation and unstable domain walls,''
  Phys.\ Lett.\  {\bf B486}, 300-305 (2000),
  [hep-ph/0002194];
  T.~Matsuda,
  ``On the cosmological domain wall problem in supersymmetric models,''
  Phys.\ Lett.\  {\bf B436}, 264-268 (1998),
  [hep-ph/9804409].
\bibitem{brane-wall}
  T.~Matsuda,
  ``Incidental brane defects,''
  JHEP {\bf 0309}, 064 (2003).
  [hep-th/0309266]:
  T.~Matsuda,
  ``String production after angled brane inflation,''
  Phys.\ Rev.\  {\bf D70}, 023502 (2004).
  [hep-ph/0403092]:
  T.~Matsuda,
  ``Formation of monopoles and domain walls after brane inflation,''
  JHEP {\bf 0410}, 042 (2004).
  [hep-ph/0406064]:
 T.~Matsuda,
  ``Formation of cosmological brane defects,''
  JHEP {\bf 0411}, 039 (2004).
  [hep-ph/0402232].
\bibitem{matsuda-necklaces}
T.~Matsuda,
  ``Dark matter production from cosmic necklaces,''
  JCAP {\bf 0604}, 005 (2006).
  [hep-ph/0509064]:
  T.~Matsuda,
  ``Topological curvatons,''
  Phys.\ Rev.\  {\bf D72}, 123508 (2005).
  [hep-ph/0509063]:
 T.~Matsuda,
  ``Primordial black holes from cosmic necklaces,''
  JHEP {\bf 0604}, 017 (2006).
  [hep-ph/0509062]:
  T.~Matsuda,
  ``Brane necklaces and brane coils,''
  JHEP {\bf 0505}, 015 (2005).
  [hep-ph/0412290].
\bibitem{Casimir-KK}
  L.~Perivolaropoulos,
  ``Vacuum energy, the cosmological constant, and compact extra dimensions: Constraints from Casimir effect experiments,''
  Phys.\ Rev.\  {\bf D77}, 107301 (2008).
  [arXiv:0802.1531 [astro-ph]];
  H.~Alnes, F.~Ravndal, I.~K.~Wehus, K.~Olaussen,
  ``Electromagnetic Casimir energy with extra dimensions,''
  Phys.\ Rev.\  {\bf D74}, 105017 (2006).
  [quant-ph/0607081];
 E.~Elizalde, S.~D.~Odintsov, A.~A.~Saharian,
  ``Repulsive Casimir effect from extra dimensions and Robin boundary conditions: From branes to pistons,''
  Phys.\ Rev.\  {\bf D79}, 065023 (2009).
  [arXiv:0902.0717 [hep-th]];
\bibitem{Casimir-KK-uni}
K.~Poppenhaeger, S.~Hossenfelder, S.~Hofmann, M.~Bleicher,
  ``The Casimir effect in the presence of compactified universal extra dimensions,''
  Phys.\ Lett.\  {\bf B582}, 1-5 (2004).
  [hep-th/0309066];
  H.~Cheng,
  ``On the Casimir effect for parallel plates in the spacetime with one extra compactified dimension,''
  Mod.\ Phys.\ Lett.\  {\bf A21}, 1957-1963 (2006).
  [hep-th/0609057].
\bibitem{Casimir-KK-T}
  L.~P.~Teo,
  ``Casimir Effect in Spacetime with Extra Dimensions: From Kaluza-Klein to
  Randall-Sundrum Models,''
  Phys.\ Lett.\  B {\bf 682}, 259 (2009)
  [arXiv:0907.2989 [hep-th]];
  R.~S.~Decca, E.~Fischbach, G.~L.~Klimchitskaya, D.~E.~Krause, D.~L.~Lopez, V.~M.~Mostepanenko,
  ``Improved tests of extra dimensional physics and thermal quantum field theory from new Casimir force measurements,''
  Phys.\ Rev.\  {\bf D68}, 116003 (2003).
  [hep-ph/0310157].
\bibitem{Casimir-KK-mixed-stable}
 A.~Chatrabhuti, P.~Patcharamaneepakorn, P.~Wongjun,
  ``AEther Field, Casimir Energy and Stabilization of The Extra Dimension,''
  JHEP {\bf 0908}, 019 (2009).
  [arXiv:0905.0328 [hep-th]].
\bibitem{Casimir-KK-warped}
R.~Linares, H.~A.~Morales-Tecotl, O.~Pedraza,
  ``Casimir force for a scalar field in warped brane worlds,''
  Phys.\ Rev.\  {\bf D77}, 066012 (2008).
  [arXiv:0712.3963 [hep-ph]].
\end{thebibliography}
\end{document}